%% file: actionability_ieee.tex
\documentclass[10pt, conference, compsocconf]{IEEEtran}
\usepackage{ifpdf}
\ifCLASSINFOpdf
\else
\fi

\usepackage{graphicx}
\usepackage{mathtools}
\usepackage{hhline}
\usepackage[utf8]{inputenc}
\usepackage{etoolbox}

\usepackage{amsmath,amssymb,amsfonts}

\usepackage{url}
\usepackage{verbatim}
\usepackage{multirow}
\usepackage{caption}
\usepackage{subcaption}

\usepackage[linesnumbered,ruled,vlined]{algorithm2e}
\usepackage{algpseudocode}
\usepackage{mathtools}

\usepackage{ifpdf}
\usepackage{lmodern}
\usepackage{lipsum}
\usepackage[T1]{fontenc}
\usepackage{textcomp}
\usepackage{mdwlist}
\usepackage{pbox}
\usepackage{comment}
\usepackage{ifthen}
\usepackage{color}
\usepackage{colortbl,xcolor}
\usepackage{upgreek}
\usepackage{upgreek}
\usepackage{todonotes}
\usepackage{wasysym}

\usepackage{tikz}
\usetikzlibrary{arrows}

\usepackage[hidelinks,bookmarks=false]{hyperref}

\definecolor{Gray}{gray}{0.9}
\definecolor{LightCyan}{rgb}{0.88,1,1}

\definecolor{LightCyan}{rgb}{0.88,1,1}
\definecolor{Gray}{gray}{0.9}
\definecolor{light-gray}{gray}{0.95}

\begin{document}
\title{Identifying Actionable Messages on Social Media}


\author{\IEEEauthorblockN{
Nemanja Spasojevic and
Adithya Rao}
\IEEEauthorblockA{Lithium Technologies | Klout\\
San Francisco, CA\\
Email: nemanja, adithya@klout.com}
}

\maketitle

\begin{abstract}
\input{texfiles/actionability_abstract}
\end{abstract}


% \fi
\IEEEpeerreviewmaketitle

\section{Introduction}
\label{section:introduction}
\input{texfiles/actionability_introduction}

\section{Related Work}
\label{section:related}
\input{texfiles/actionability_related_work}

\section{Methodology}
\label{section:model_generation}
\input{texfiles/actionability_model_generation}

\section{System Overview}
\label{section:system_overview}
\input{texfiles/actionability_system_overview}

\section{Lexicon Generation}
\label{section:lexicons}
\input{texfiles/actionability_lexicons}

\section{Feature Generation}
\label{section:features}
\input{texfiles/actionability_features}

\section{Evaluation and Results}
\label{section:model_evaluation}
\input{texfiles/actionability_model_evaluation}

\section{Conclusion}
\label{section:conclusion}
\input{texfiles/actionability_conclusion}

\section{Acknowledgment}
\label{section:conclusion}
\input{texfiles/actionability_acknowledgement}

\bibliographystyle{IEEEtran}
\bibliography{actionability_ieee}

\end{document}

%% file: texfiles/actionability_abstract.tex
Text actionability detection is the problem of classifying user authored natural language text, according to whether it can be acted upon by a responding agent.
In this paper, we propose a supervised learning framework for domain-aware, large-scale actionability classification of social media messages. 
We derive lexicons, perform an in-depth analysis for over 25 text based features, and explore strategies to handle domains that have limited training data.
We apply these methods to over 46 million messages spanning 75 companies and 35 languages, from both Facebook and Twitter.
The models achieve an aggregate population-weighted F measure of 0.78 and accuracy of  0.74, with values of over 0.9 in some cases.

%% file: texfiles/actionability_introduction.tex
Social media has emerged as a public utility where users can post messages to receive responses. 
% Some of these messages such as Facebook interactions with friends are private in nature, while others such as Twitter messages and public posts on Facebook Pages are intended to be broadcasted to all network users. 
Because of the democratic nature of public interactions on social networks, they provide an ideal platform to interact with brands, companies and service providers. 
According to Okeleke \cite{ovum2014white} $50\%$ of users prefer reaching service providers on social media over contacting a call center.
On the other side of these interactions, many companies are developing strategies to respond to their customers on social networks.

% Over the past decade hundreds\footnote{https://hootsuite.com}\footnote{https://tweetdeck.twitter.com}\footnote{https://www.sprinklr.com} of social media tools have emerged to help brands and power users monitor their accounts, or aggregate sentiment towards them.
% Because of this, sentiment analysis on social networks is a well-studied problem.
Over time, social media tools have been used for analyzing user interests \cite{nemanja-lasta}, broadcasting marketing messages to maximize responses \cite{Spasojevic:when-to-post}, and most recently for managing individual conversations.
For such conversations, social media messages that include a clear call to action, or raise a specific issue can be identified by an agent of the company, who may provide a helpful response.
Such messages can be categorized as \textit{actionable}.
Alternatively, agents may not be able to respond to messages that are too broad, general or not related to any specific issue.
These messages may be categorized as being \textit{non-actionable}.

The ability to sift through interactions and classify them as actionable can help reduce company response latency and improve efficiency, thereby leading to better customer service.
A good solution to identify actionable messages can lead to large cost savings for companies by reducing call center loads. 
Here we propose a framework to solve this under-explored problem at scale.

The difficulty in identifying actionability arises from factors such as scarcity of actionable messages when dealing with large volumes of messages, diversity of languages spoken by customers and subtle differences in Twitter and Facebook natural language patterns. 
Further, depending on the posting intention of the user and the company under question, an agent may consider a message actionable under varying contexts such as customer support, product complaints, sales, or engagement opportunities.

Our contributions in this study are:
\begin{itemize}
  \item \textbf{Lexicon and Feature Generation:} We introduce methods for actionability lexicon generation, and explore more than 25 features to classify messages.
  \item \textbf{Scale:} We propose a framework that scales across over 75 companies, 35 languages, 2 sources, generating over 900 domain models for over 46 million messages.
  \item \textbf{Optimal Domain Selection:} For domains with limited training data, we describe strategies that leverage overlapping attributes from the entire dataset.
\end{itemize}

%% file: texfiles/actionability_related_work.tex
The problem of actionability has been studied before by Munro \cite{actionableHaiti} in the context of the disaster relief, where subword and spatiotemporal models were used to classify short-form messages. 
The dataset used contained more than $100,000$ messages in English and French which originated during the 2010 Haiti earthquake, sourced from Twitter, SMS, and radio.
The F measure on only word and n-gram based features was $0.33$, while the final F measure was $0.885$ when geographic location-based features were included. 
This study showed that relying simply on textual features may be insufficient for identifying actionability.

More research has been done in the context of customer complaint identification. 
Jin et. al. \cite{jin2013service} used TF-IDF derived from text, along with KNN and SVM techniques, to identify complaints with an F measure of $0.8$ on a dataset containing $5,500$ messages from an online Chinese hotel community.
Although the study showed promising results the dataset was limited and constrained to a very specific domain.

A significant body of work regarding the text classification on social media has focused on the sentiment analysis. 
A wide range of research has been carried out, deploying different techniques such as
part-of-speech tagging \cite{agarwal2011sentiment}, lexicon approaches \cite{taboada2011lexicon}, deep learning \cite{glorot2011domain}, \cite{socher2013recursive}, and hybrid approaches \cite{khan2014tom}, \cite{Jin:2009:ONM:1557019.1557148}.
Human disagreements on the judgement of sentiment poses a challenge, limiting models to have precision and recall values in the range of $0.75-0.82$ \cite{nigam2004towards}.

While the majority of research investigates text classification problems within a unified domain, it has been recognized that sentiment detection models may change depending on the domain context \cite{kanayama2006fully}, \cite{glorot2011domain}.
Specific domain adaptation techniques have been shown to improve performance compared to generalized domain techniques.
Glorot et al. \cite{glorot2011domain} were able to significantly improve quality over the baseline using domain adaptive models applied on $22$ different domains.
Hiroshi et al. \cite{kanayama2006fully} used a lexicon approach that achieved precision values of more than $0.9$ on $4$ different domains.
In our study we consider over $900$ domains.

%% file: texfiles/actionability_model_generation.tex
\subsection{Domain Representation}

Each domain under consideration may have the distinct attributes described below:
\begin{itemize}
   \item \textbf{Company ($c$)}: Company mentioned or associated with the terms in the message.
   \item \textbf{Language ($l$)}: Language of the message.
   \item \textbf{Source ($s$)}: The social network on which the message originated.
\end{itemize}

A fully specified domain $D$ can be represented as a set $\{c,l,s\}$. 
For such a domain, we consider the associated domains given by the elements of the power set of $D$:

\begin{small}
  \vspace{-1.0\baselineskip}
  \[ \mathcal{P}(D) = \{\{\}, \{c\}, \{l\}, \{s\}, \{c,l\}, \{l,s\}, \{c,s\}, \{c,l,s\}\} \]
  \vspace{-1.0\baselineskip}
\end{small}

% \begin{equation}
%   \tiny
%   \mathcal{P}(D) = \{\{\}, \{c\}, \{l\}, \{s\}, \{c,l\},
%                      \{l,s\}, \{c,s\}, \{c,l,s\}\} \nonumber
% \end{equation}

Each partially specified member of $\mathcal{P}(D)$ denotes a generalized domain spanning multiple fully specified ones. 
For example, the domain given by \{`nokia', `es'\} pertains to the domain consisting of messages in Spanish that were associated with Nokia across all sources under consideration.
Thus each fully specified domain has 7 other generalized domains.

\subsection{Problem Statement}

Let us consider a specific domain $D$, for which we want to identify actionable messages.
Let $M_D$ be the set of messages, known to be created under $D$, and
labeled as actionable or non-actionable.
Our goal is to be able to classify any new messages created under $D$ and
assign them the appropriate labels.

Note that we operate within the scenario where the domain $D$ of the message is known a priori, or derived before classification.
We are therefore not trying to perform multi-class classification of messages across multiple unknown domains, but instead aim to perform binary classification within many known and disjoint domains. 
Thus this is a different problem than the one considered by typical domain adaptive techniques.

\subsection{Classification}

For each labeled data point $m \in M_D$, we derive features that capture different aspects of actionability.
Thus each message is associated with a feature vector $\textbf{f}(m)$ and a binary class label $a(m) \in \{ `actionable', `non-actionable'\} $.
We describe the derivation of these features in more detail in Section \ref{section:features}.

We divide the labeled set $M_D$ into a training set $T_D$ and an evaluation set $E_D$.
We train classifiers on $T_D$ using multiple supervised learning techniques, and the best classifier is picked based on a $F$ measure evaluated using cross-validation on $T_D$. 
The final results presented are evaluated on the held out set $E_D$.
The techniques used for classification are described in more detail in Section \ref{section:model_evaluation}.

\subsection{Model Selection}

One of the challenges in the problem described above is that the data available for a fully specified domain may sometimes be insufficient for the purposes of training a good classifier.
For example, if the number of messages associated with the domain 
$\{`nokia',`es',`tw'\}$ is small, then this scarcity may not allow a good model to be learnt.
However, the domain given by $\{`nokia', `tw'\}$, which spans messages for Nokia across multiple languages on Twitter, may have more data points. 
This generalized domain may  yield a better classifier that to be applied to the original domain.

Let $D$ be a domain that is fully specified by a company, language and source. The best applicable domain for $D$ is chosen in the following manner.
For each domain $D^*$ in $\mathcal{P}(D)$, we aggregate messages and split them into a training set $T_{D^*}$ and an evaluation set $E_{D^*}$.
We derive a set of classifier models $\mathcal{C}_{D^*}$ for each $D^*$ in $\mathcal{P}(D)$.
Let $\mathcal{C}_D$ be the union of these sets, given by $\mathcal{C}_D = \bigcup_{D^* \in \mathcal{P}(D)} \mathcal{C}_{D^*}$.
Each classifier model in $\mathcal{C}_D$ is evaluated with cross-validation on $T_{D}$ and the best classifier is chosen, based on an $F$ measure.

Using this methodology, we build $48,384$ distinct models on $1,778$ domains, of which $937$ are fully specified.
We further explore the tradeoff between the number of training samples and domain specificity in Section \ref{section:model_evaluation}.

%% file: texfiles/actionability_system_overview.tex
\subsection{Datasets}

% Avaiable here
% https://docs.google.com/a/klout.com/drawings/d/1FE1nE3oJVnAWJb3YXuw3PpihhOfE4g31rGmmoedkMHg/edit
The research undertaken in this paper was performed on a social media management platform that helps agents respond to customer posts on behalf of their company.
The platform prioritizes incoming messages from customers and routes them to the appropriate agents, and is integrated with major social networks, brand communities,
and online forums. 
Thus agents using the platform may respond to messages that are actionable, and ignore those that are not.
If an agent provided a response, then the message is marked \textit{actionable}, otherwise it is marked \textit{non-actionable}.
This labeled dataset is then used for training and evaluating models.

In this study we used a trailing window of 6 months of data, from November 1st 2014 to May 1st 2015. 
This included $46$ million unique messages, of which $0.7$ million were found to be actionable by agents. 
We pre-processed the dataset so that the labeled set is balanced, with $50\%$ positive and $50\%$ negative examples under each fully specified domain. 
All data analyzed here is publicly available on Twitter and Facebook and no private data was used as part of this research.

\subsection{System Components}

\begin{figure}
  \centering
  \fbox{\includegraphics[width=0.97\columnwidth]{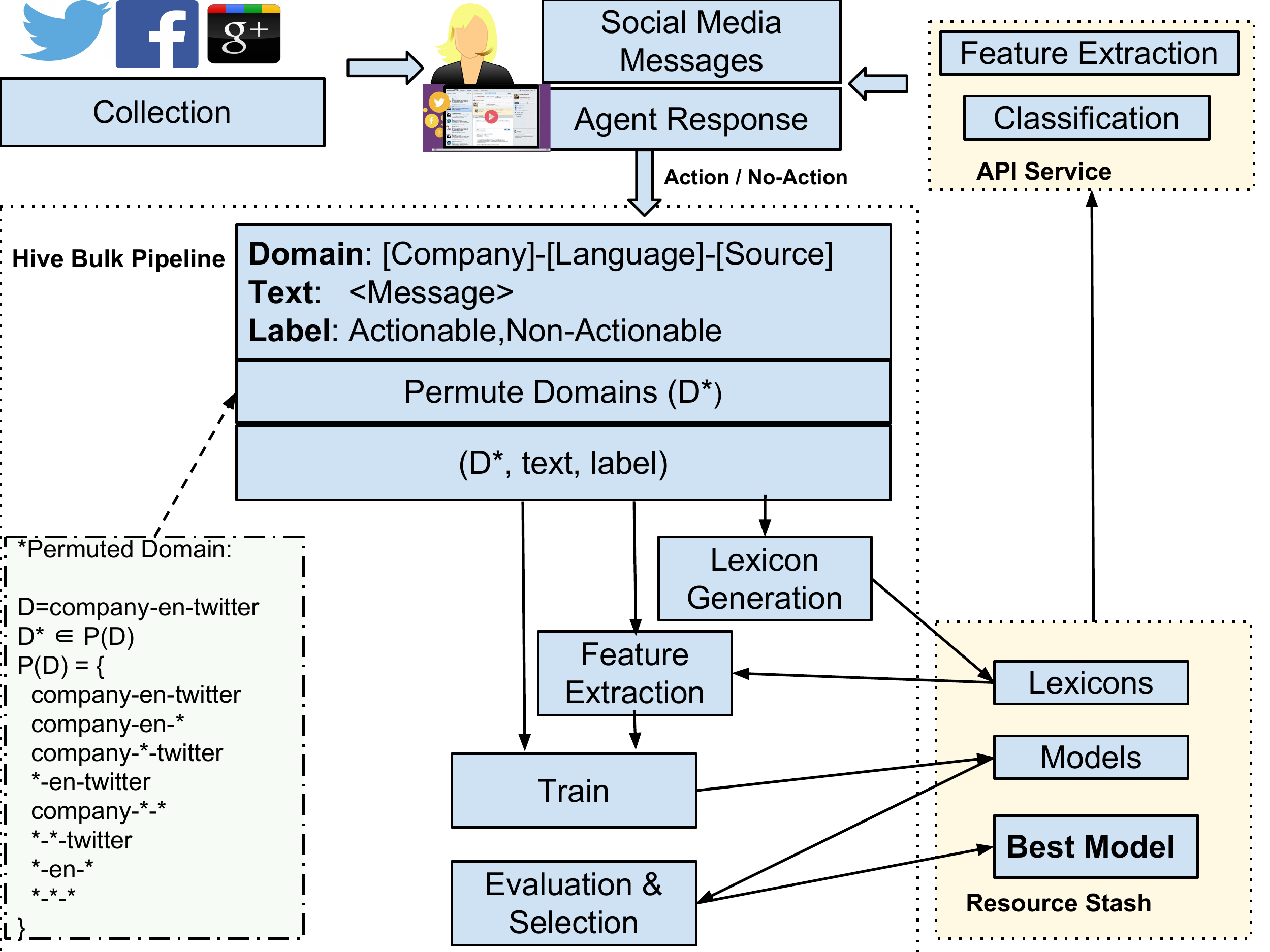}}
  \setlength{\abovecaptionskip}{0pt}
  \caption{System Overview}
  \label{figure:system_overview}
  \vspace{-0.12in}
\end{figure}

The system overview is shown in Figure \ref{figure:system_overview}.
Data is ingested into the platform from social networks such as Twitter, Facebook and Google+, and messages are routed to company agents based on the mentioned company.
The agent responses to the messages are recorded for each domain, which serves as the labeled data for model training and evaluation. 
Lexicon dictionaries are built for each permuted domain, and features are extracted from the message text.
Models are trained using the features and labels, and are evaluated on a held out dataset. 
The best model is chosen for each domain, which is then used to classify new incoming messages as actionable or non-actionable.

The collected public engagement data is written out to a Hadoop cluster that uses HDFS as the file system.
We implement independent Java utilities with Hive UDF (User Defined Function) wrappers. These utilities include functions to perform feature extraction and derive generalized domains, among other functions.
The combination of Hive Query Language (HQL) and UDFs allows to quickly build map-reduce jobs, where trivial data transformations are performed using HQL markup, while complex logic and functionality is abstracted to the UDFs.
For machine learning we use Hivemall \cite{Yui15}, \cite{bigdata13myui}, a scalable  library designed specifically for Hive.

The data processing pipeline that generates the lexicons and models is deployed in production and runs on a daily basis, using a 6 month trailing window of data.
On a 150-node cluster, this pipeline has a cumulative daily I/O footprint of 73GB of reads, 22GB of writes, and 4.95 days of CPU usage. 

%% file: texfiles/actionability_lexicons.tex
In order to identify keywords that hold information about actionability and sentiment, we first generate lexicons for each domain.

\paragraph{Actionability Lexicons}
Actionable messages may contain specific keywords depending on the context, intention, and the expected reaction from the responding agent.
To account for context specific keywords, we build lexicons for each domain and label class.
The message text is tokenized using the Apache Lucene \footnote{https://lucene.apache.org/} tokenizer since it handles a diverse set of languages.
Keyword scores are derived for each term in the corpus using a variation of the document frequency measure.
The final lexicons are represented as word vectors for each domain-class label pair. 

Let $d$ be a particular domain and $a$ be the actionability class label in $\{'actionable', 'non\_actionable'\}$. Then for each domain-class pair $(d,a)$, we first compute the normalized document frequency for each term $t$ as: 
\begin{equation}
 \small
  ndf^{d,a}(t) = \frac{n^{d,a}(t)}{|M_{d,a}|} \\
\end{equation}
where $M_{d,a}$ is the corpus of all documents (messages) for $(d,a)$. 
To eliminate the effect of outliers, we adjust the above quantity with the $95^{th}$ percentile value computed on the set of all terms $T_{d,a}$ for $(d,a)$:
\begin{equation}
 \small
  adf^{d,a}(t) = \frac{log(ndf^{d,a}(t))}{log(percentile_{0.95}(ndf^{d,a}_{t \in T_{d,a}}(t))} \label{eq:document_frequency}  \\
\end{equation}
This adjustment ensures that different lexicons are comparable across domains. 

\begin{figure}
\centering
% Shift globally figure up
\begin{subfigure}[b]{0.235\textwidth}
  \centering
  \includegraphics[width=1.0\textwidth]{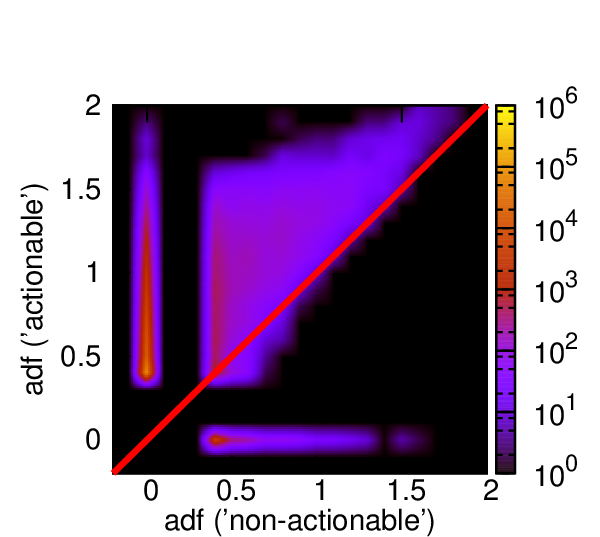}
  \vspace{-0.3\baselineskip}
  \caption{\{`yahoo',`en',`tw'\}}
  \label{fig:2d_dictionary_british_gas_na_na}
\end{subfigure}
\begin{subfigure}[b]{0.235\textwidth}
  \centering
  \includegraphics[width=1.0\textwidth]{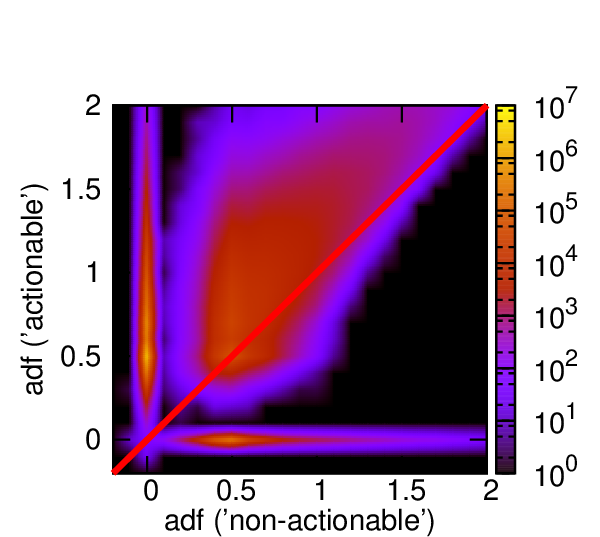}
  \vspace{-0.3\baselineskip}
  \caption{\{`en', `tw'\}}
  \label{fig:2d_dictionary_global_na_na}
\end{subfigure}

\caption{Cross-label distribution of adjusted document frequency ($adf$)}
\label{fig:model_performance_by_type}
\end{figure}

The adjusted document frequency distributions across actionable and non-actionable labels for the domains of \{`yahoo',`en',`tw'\} and \{`en',`tw'\} are shown in Figure \ref{fig:model_performance_by_type}.
By observing the keyword distributions we conclude that most keywords are indicative of non-actionable content, and only a small set of keywords just below the diagonal represent actionable content.
Ambivalent keywords sit on the diagonal, while highly polarizing keywords belonging to a single lexicon are pushed towards the x and y axes.
Thus we see that actionable and non-actionable messages have significant overlap in the keyword space, yet display polarizing keywords that can separate the classes.

While the adjusted document frequency quantity may itself be used to create the lexicons, we found in our experiments that including ambivalent keywords that appear in both class lexicons for a domain leads to poorer results.
We therefore further constrain each keyword to appear in only one class lexicon per domain.
Thus we compute the keyword score $w_{d,a}(t)$ for the term $t$ as follows:
\begin{equation}
 \small
  w^{d,a}(t) = \begin{cases} \label{eq:dictionary_vector}
        adf^{d,a}(t) - adf^{d,\hat{a}}(t), & \text{if}\ adf^{d,a}(t) > adf^{d,\hat{a}}(t)\\
        0                                , & \text{otherwise} \\
    \end{cases}
\end{equation}
This approach led to about a $1.5\%$ increase in the F measures for the final models.

Table \ref{table:dictionary_examples} shows the top keywords for both the adjusted document frequency and the constrained keyword score for the domain of \{`yahoo', `en', `tw'\}.
The adjusted document frequency yields actionable words such as \textit{`help', `mail', `can'}, and for non-actionable ones like \textit{`via', `yahoofinance', `yahoofantasy'}.
Note that some keywords are present in both labels, but the relative order still captures the importance with respect to each label.

The actionable keywords from the constrained word scores $w$ are more strongly indicative of context, than the ones from adjusted document frequency.
Another point to note is that many keywords are very specific to the domain (in this case services of the company `Yahoo!'), highlighting the importance of domain specificity.
% yahoo!
% container02.prodlsw:7-en-twitter
% actionable_score
% {"i":1.9692124554012247,"my":1.9360347154358362,"to":1.9328672450801816,"the":1.8914914883188223,"is":1.8469229816015387,"a":1.8249249035667912,"it":1.811533207150765,"you":1.7946324998082115,"help":1.7889434722534239,"in":1.7780139756561646,"on":1.7744655867921562,"have":1.7675961324288387,"me":1.766435660417677,"can":1.7537859854237718,"not":1.7508335072082648,"mail":1.7232189978255024,"this":1.7216535349421391,"of":1.718825489303536,"can't":1.7186020515972167,"that":1.717266184140734}
% non_actionable_score
% {"via":2.2717391415231676,"yahoofinance":2.2159935443010577,"the":2.2124895587026057,"to":2.2108618192506646,"a":2.145520844900698,"in":2.1407728246086934,"of":2.121601373995616,"i":2.108410754370511,"is":2.1062554394668016,"on":2.094564671861213,"yahoofantasy":2.082344878159879,"you":2.0591167052318347,"my":2.0550767482691557,"or":2.0390764517103808,"with":2.0296450164299387,"your":2.028024123059031,"it":2.0228797563805796,"this":2.0186589728592303,"at":2.0024518965314178,"that":1.9900129317117057}
\begin{table*}
  \small
  \begin{center}
    \caption{Adjusted Document Frequency and Lexicon Top Keywords For \{`yahoo', `en', `twitter'\}. }
    \begin{tabular}{ | l | l | p{11cm} |}
    \hline
    \textbf{Metric}               & \textbf{Label}    & \textbf{Top Keywords}  \\ \hline
    \multirow{2}{*}{$adf$ }       & actionable            & i,my,to,the,is,a,it,you,help,in,on,have,me,can,not,mail,this,of,can't,that \\\cline{2-3}
                                  & non\_actionable   & via,yahoofinance,the,to,a,in,of,i,is,on,yahoofantasy,you,my,or,with,your,it \\ \hline
    \multirow{2}{*}{$w$}          & actionable            & password,can't,or,still,back,app,change,sent,aviate,locked,issue,changed,keep \\\cline{2-3}
                                  & non\_actionable   & out,yahoonews,yahoomail,ysmallbusiness,small,obama,homescreen,report \\ \hline
    \end{tabular}
  \end{center}
  \label{table:dictionary_examples}
\end{table*}

\paragraph{Sentiment Lexicons}

Actionable messages usually contain complaints from users about services or products provided by the brand.
Negative sentiment is prominent in user complaints, but not all messages with negative sentiment will be actionable. 
To capture this relationship between sentiment and actionability, we introduce positive and negative keyword lexicons as well. 
The word weights are derived from SentiWordNet 3.0 \cite{baccianella2010sentiwordnet}, and the value assigned to each keyword is the sum of all SentiWebNet $PosScore$, or $NegScore$, ignoring the keyword's part-of-speech tag.
Unlike the previously discussed actionability lexicons, sentiment lexicons are available only for English. 
Incorporating sentiment lexicons for other languages is left for future work.

%% file: texfiles/actionability_features.tex
Analysis of social media text is a well studied problem. 
Different feature extraction approaches that encapsulate the length of the message, the presence of hashtags, mentions, question marks, sentiment and topical content have been explored. 
In addition to content-based features, many studies have leveraged users' demographic information, as well as features derived from the network topology. 
In our research we focus primarily on content-based features, since the goal is to predict if a specific message is actionable, irrespective of the user posting it.

\paragraph{Lexicon Features}

As explained above, we generate actionable and non-actionable lexicons for each domain-label pair, and positive and negative sentiment lexicons for English using SentiWordNet.
The lexicons are represented as word vectors, with the keyword scores as elements of the vector.
The lexicon-based features for a message are then derived as the dot products of the term frequency vector and the lexicon word vectors, scaled by total number of words in the message:

{ \small \[ f_k(d) = \frac{\vec{tf}(d) \cdot \vec{w}}{|d|} \] }
Since the lexicon vectors are designed to be orthogonal, these features are valuable to achieve separation between the classes.

\paragraph{Marker Features}

We also derive a set of features from markers in the text.
Markers can be special characters, or words having special meanings.
\begin{itemize}
\item \textbf{`?', `!':} Exclamation marks are frequently used with actionable messages to add emphasis to the content, while question marks are clear indicators of questions or confusion.
\item \textbf{`via', `rt':} Markers like `rt', and `via' have emerged in social media as a way to credit original content authors. Capturing these markers helps identify if the user is the original creator of the message.
\item \textbf{`@':} Input text on social media may often include tags or mentions of other users. This is done by prefixing user account names with `@'. Such markers are used to draw attention to the content by claiming a relation between the message and the mentioned user.
\item \textbf{`\#':} Hashtags preceding a keyword marks content with specific topics. They are good indicators that the message has certain topics associated with it.
\item \textbf{`url':} URL link presence in a message may indicate actionability in certain contexts, and non-actionability in others.
\end{itemize}

For each occurrence of the marker we generate a separate feature, by appending the occurrence count to the marker name (eg. $@-0$, $\#-1$ etc).
For a marker $ch$ within document $d$, an $index$ function returns the position of marker within the document. 
The feature value for $i$ occurrences is then calculated as:

{ \small
  \[
  f^{i}_k(d) =
  \begin{cases}
      1 - \frac{index(ch, d)}{length(d)},& \text{if } ch \in \{`@'\} \\
      \frac{index(ch, d)}{length(d)}    ,& \text{otherwise}
  \end{cases}
  \]
}

By capturing the occurrence count as well as the position of the marker, we incorporate more information into these features.

\paragraph{Readability Features}

In order to examine the relationship between the readability of a message and its actionability, we generate features that measure the former. 
Readability scores are schemes that try to measure how difficult it is to comprehend a given document.
In particular we use Dale-Chall \cite{dale1948formula} and Flesch-Kincaid readability scores \cite{kincaid1975derivation}.
We use these features for English messages only. 

The Dale-Chall readability score depends on a manually curated list of about $3,000$ simple words. 
For each document $d$, a formula is then derived using the total number of sentences $S(d)$, words $W(d)$, and difficult words $W_d(d)$:
\begin{equation}
\small
  f_k(d) = 0.159  \times \frac{W_d(d)}{W(d)} + 0.0496 \times \frac{W(d)}{S(d)}
\end{equation}
The Flesch–Kincaid readability scheme further use syllables $s(d)$, to compute a score:
\begin{equation}
\small
  f_k(d) = 0.39 \times \frac{W(d)}{S(d)} + 11.8 \times \frac{s(d)}{W(d)} - 15.59
\end{equation}
Dale-Chall scores range from 4.9 for text easily understood by a 
$4^{th}$ grader to $10$ and above for text easily understood by graduate students.
Flesch-Kincaid readability scores range from $0$ to $100$, with a higher score indicating greater ease of readability. 
For our purposes, both scores are rescaled to a $[0, 1]$ range to create readability features for messages.

\paragraph{Emoticon Features}

Emoticons are pictorial representations of emotions widely used in social media to augment textual content with mood and tone. 
Variations of emoticons include kaomoji -- horizontal Japanese style emoticons (eg. `(*\_*)') -- and emoji -- Japanese ideograms (eg. `\smiley') captured by the unicode standard.\footnote{http://www.unicode.org/}
Here we refer to emoticons, kaomoji, and emoji as simply emoticons.

Using emoticons for inference of text sentiment is a well-studied problem \cite{zhao:2012:mes:2339530.2339772}, \cite{kiritchenko2014sentiment}. 
In our case we map the emoticons to their English language descriptions, which are then mapped to sentiment using the sentiment lexicons described above. 
For emoji we used descriptions provided by the unicode consortium.
Emoticons and kaomojis were mapped to descriptions using public domain catalogs
\footnote{http://cool-smileys.com/text-emoticons} containing over $1,300$ emoticons.
For unmapped emoticons, we derive polarity using Twitter's `EmoticonExtractor' as the fallback strategy.
The presence of emoticons in a message, and the scaled amplitudes of positive or negative sentiment, are derived as features.

\paragraph{Document Length}

As the text length is correlated with information quantity in the text, we capture document length feature values as: $ f_{k}(d) = \frac{|d|}{N}, N \in \{100, 1000\}$.

\subsection{Feature Evaluation}

Table \ref{table:features} shows the Mutual Information (MI) and coverage for the features described above. 
A higher MI value indicates that the feature contains information that is more useful for making predictions. 
We find that the features derived from the actionable lexicon have one of the highest MI values, validating the use of these domain-specific lexicons. 
The highest coverage is seen for the non-actionable lexicon features, highlighting that most messages contain non-actionable text.

Sentiment lexicon-based features seem to be poor predictors, perhaps because the words used in these lexicons are not typically indicative of actionability. 
That, however, does not imply that sentiment itself is a poor predictor. 
Its importance is evident from the feature for the presence of a negative emoticon, which has the highest MI value among all. 

\begin{table}[ht]
\small
\setlength\tabcolsep{4pt}
\caption{Mutual Information and Coverage}
\begin{tabular}{|l||r|r|}
  \hline
  \textbf{Feature} & \parbox[t]{1.9cm}{\textbf{Mutual \newline Information $[ 10^{-3} ]$}} & \parbox[t]{1.9cm}{\textbf{Coverage}} \\
  \hline
  \multicolumn{3}{|c|}{\textbf{Lexicon Features}}  \\ \hline
  actionable     & 38.332       & 67.35\%  \\ \hline
  non actionable & 3.382        & 87.21\%     \\ \hline
  negative       & 0.117        & 27.63\%  \\ \hline
  positive       & 0.067        & 27.96\%  \\ \hline

  \multicolumn{3}{|c|}{\textbf{Marker Features }}  \\ \hline
  ?-0     & 16.365 & 18.82\% \\ \hline
  ?-1     & 2.572  &  3.96\% \\ \hline

  !-0     & 0.296  & 14.01\% \\ \hline
  !-1     & 0.245  &  5.58\% \\ \hline

  rt-0    & 5.268  & 0.73\%  \\ \hline
  rt-1    & 0.098  & 0.02\%  \\ \hline

  via-0   & 6.266  & 0.71\%  \\ \hline
  via-1   & 0.004  & 0.00\%  \\ \hline     % 0.0006

  @-0     & 0.927  & 39.05\% \\ \hline
  @-1     & 14.840 &  8.70\% \\ \hline

  \#-0    & 29.833 & 10.98\% \\ \hline
  \#-1    & 16.282 &  4.24\% \\ \hline

  has-url & 35.533 & 17.73\% \\ \hline

   \multicolumn{3}{|c|}{\textbf{Readability Features}}  \\ \hline

  Dale-Chall     & 0.036 & 25.46\% \\ \hline
  Flesch-Kincaid & 0.039 & 28.80\% \\ \hline

   \multicolumn{3}{|c|}{\textbf{Emoticon Features}}  \\ \hline
  has any emoticon       & 21.006  & 28.18\% \\ \hline
  negative emoticon      & 39.188  & 17.87\% \\ \hline
  positive emoticon      & 0.713   &  3.93\%  \\ \hline

   \multicolumn{3}{|c|}{\textbf{Length Features}}  \\ \hline

  > 10 characters   & 0.173    & 55.19\% \\ \hline
  > 100 characters  & 0.421    & 42.47\% \\ \hline
  > 10 words        & 4.091    & 62.30\% \\ \hline
  > 100 words       & 0.003    &  0.01\% \\ \hline % 0.0126

\end{tabular}
\label{table:features}
\end{table}

%% file: texfiles/actionability_model_evaluation.tex
We employ supervised learning to build models for each domain, using the feature vectors generated for the messages and the associated labels. 
We evaluate the performance of the models, and compare results across different domain specificities, networks, languages and companies.
We also provide examples of specific messages and their corresponding labels as determined by the models.

\begin{figure*}
\centering
% Shift globally figure up
\vspace{-1.7\baselineskip}

  \begin{subfigure}[b]{0.32\textwidth}
    \centering
    \includegraphics[width=0.95\textwidth]{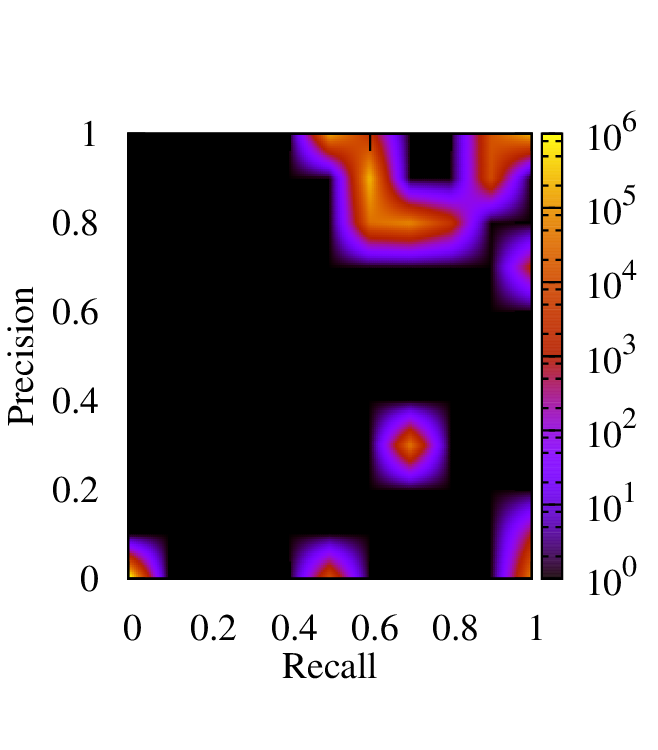}
    \vspace{-1\baselineskip}
    \caption{Logistic model for \{c, l, s\} \newline domain (strategy \textbf{A})}
    \label{fig:population_count_logistic-full_spec_prec_rec_2d}
  \end{subfigure}
  %
  % B - Logistic model for {} domain ??
  %
  \begin{subfigure}[b]{0.32\textwidth}
    \centering
    \includegraphics[width=0.95\textwidth]{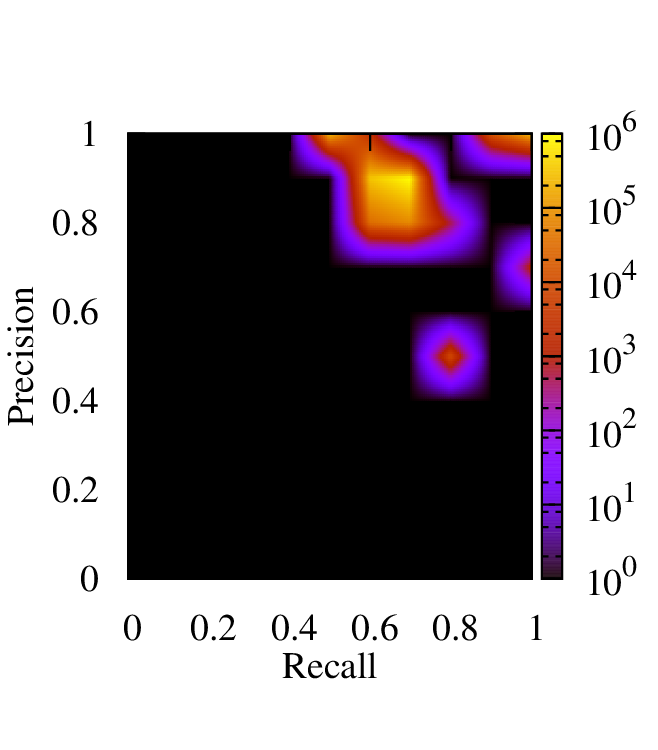}
    \vspace{-1\baselineskip}
    \caption{Logistic for best domain \newline (strategy \textbf{C})}
    \label{fig:population_count_logistic-best_permuted_prec_rec_2d}
  \end{subfigure}
  \begin{subfigure}[b]{0.32\textwidth}
    \centering
    \includegraphics[width=0.95\textwidth]{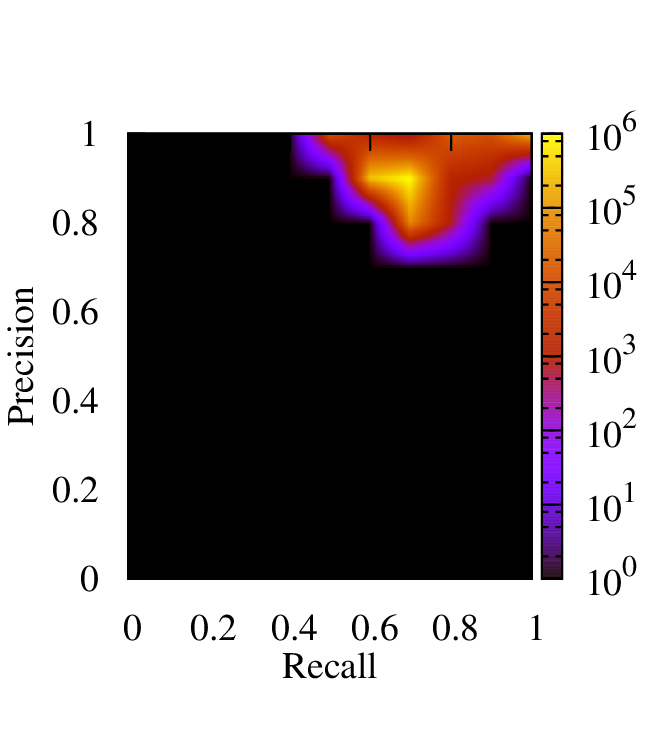}
    \vspace{-1\baselineskip}
    \caption{Best model-domain \newline (strategy \textbf{D})}
    \label{fig:population_count_best_model-best_permuted_prec_rec_2d}
  \end{subfigure}

\caption{Precision and recall distributions for different company, language, source domains}
\label{fig:prec_rec_2d_distr}
\end{figure*}

\paragraph{Specificity and Strategy}
As described before, the dataset for a fully specified domain may not always be sufficient to build good models, and sometimes a generalized domain may yield better results. 
On the other hand, a domain that is too general may not have enough contextual information to make accurate predictions.
In this section we examine the best domain selection from the elements of $\mathcal{P}(D)$ for each fully specified domain $D$.

Table \ref{table:domain_specificity_selection} shows the breakdown of domain selection for all the $937$ fully specified models.
% All domains were trained using logistic regression.
The fully specified domains are themselves selected in $60\%$ of the cases. 
However, these are not the best choices in $40\%$ of the cases, validating the need for generalized domains. 
In $16.7\%$ of cases the generic source models are chosen, showing that for some brands messages across Twitter and Facebook are similar, for the given language.
The most general model is selected only in $8.9\%$ of the cases, implying that a non-contextual global model has limited success in predicting actionability. 
% \mathcal{P}(D) = \{\{\}, \{c\}, \{l\}, \{s\}, \{c,l\}, \{l,s\}, \{c,s\}, \{c,l,s\}\} \nonumber

\begin{table}[ht]
\small
\caption{Domain Specificity Selection Comparison ($s$ - source, $c$ - company, $l$ - language)}
\centering
\begin{tabular}{|l||r|r|}
  \hline
  \rowcolor{Gray} \parbox[t]{2cm}{\textbf{Domain \newline Category}} & \parbox[t]{2cm}{\textbf{Times \newline Selected}} & \parbox[t]{2cm}{\textbf{Percent \newline Selected}} \\
  \hline % \hhline{|=||=|=|}
  $\{c,l,s\}$         & 566 & 60.4\%  \\ \hline % COMPANY-LANGUAGE-SOURCE
  $\{c,l\}$           & 157 & 16.7\%  \\ \hline % COMPANY-LANGUAGE-*
  $\{\}$              & 84  & 8.9\%   \\ \hline % *-*-*
  $\{c,s\}$           & 52  & 5.5\%   \\ \hline % COMPANY-*-SOURCE
  $\{l\}$             & 29  & 3.1\%   \\ \hline % *-LANGUAGE-*
  $\{c\}$             & 25  & 2.7\%   \\ \hline % COMPANY-*-*
  $\{l,s\}$           & 21  & 2.2\%   \\ \hline % *-LANGUAGE-SOURCE
  $\{s\}$             & 3   & 0.3\%   \\ \hline % *-*-SOURCE
\end{tabular}
\label{table:domain_specificity_selection}
\end{table}

\begin{table}[ht]
\small
\caption{Model Technique Selection Comparison}
\centering
\begin{tabular}{|l||r|r|}
  \hline
  \rowcolor{Gray} \parbox[t]{3cm}{\textbf{Model \newline Strategy}} & \parbox[t]{1.25cm}{\textbf{Times \newline Selected}} & \parbox[t]{1.25cm}{\textbf{Percent \newline Selected}} \\
  \hline % \hhline{|=||=|=|}
  AROW                     & 378 & 40.4\% \\ \hline
  Logistic                 & 245 & 26.2\% \\ \hline
  Soft Confidence Weighted & 89  & 9.5\%  \\ \hline
  Adagrad RDA              & 82  & 8.7\%  \\ \hline
  Passive Aggressive       & 53  & 5.6\%  \\ \hline
  Confidence Weighted      & 46  & 4.9\%  \\ \hline
  Perceptron               & 44  & 4.7\%  \\ \hline
\end{tabular}
\label{table:technique_selection}
\end{table}

For each domain, we further experimented with different supervised learning techniques, and compare their selection performance in Table \ref{table:technique_selection}. 
We observe that AROW \cite{Crammer09adaptiveregularization} is the most chosen technique, followed by Logistic Regression and Soft Confidence Weighting \cite{Wang12exactsoft}. 

% Figure \ref{fig:model_distribution_by_type} shows the namespace distributions for precision, recall, F1 and accuracy metrics as compared across the different techniques. The figure visualizes that AROW is typically chosen because it has higher values on the metrics, for more namespaces.

% \begin{figure*}
% \centering
% \begin{subfigure}[b]{0.2\textwidth}
%   \centering
%   \includegraphics[width=0.95\textwidth]{figure/eps/multi_model_prec_rec/multi_model_prec_rec_precision}
%   \caption{Precision}
%   \label{fig:model_type_prec_distr}
% \end{subfigure}
% \begin{subfigure}[b]{0.2\textwidth}
%   \centering
%   \includegraphics[width=0.95\textwidth]{figure/eps/multi_model_prec_rec/multi_model_prec_rec_recall}
%   \caption{Recall}
%   \label{fig:model_type_rec_distr}
% \end{subfigure}
% \begin{subfigure}[b]{0.2\textwidth}
%   \centering
%   \includegraphics[width=0.95\textwidth]{figure/eps/multi_model_prec_rec/multi_model_prec_rec_f1}
%   \caption{F1}
%   \label{fig:model_type_f1_distr}
% \end{subfigure}
% \begin{subfigure}[b]{0.2\textwidth}
%   \centering
%   \includegraphics[width=0.95\textwidth]{figure/eps/multi_model_prec_rec/multi_model_prec_rec_accuracy}
%   \caption{Accuracy}
%   \label{fig:model_type_acc_distr}
% \end{subfigure}
% \caption{Stat Distribution for different model types \cite{Crammer09adaptiveregularization}, \cite{Wang12exactsoft}}
% \label{fig:model_distribution_by_type}
% \end{figure*}

Next, we evaluate the optimal model for a domain via a combination of specificity and learning techniques. 
Our baseline strategies use logistic regression for the fully specified domain (\textbf{A}) and for the fully generalized one (\textbf{B}). 
We compare these baselines to a strategy \textbf{C} that uses logistic regression and picks the best domain from $\mathcal{P}(D)$, and strategy \textbf{D} that picks the best domain as well as the best learning technique. 
 
% Fourth strategy \textbf{D} uses best performing domain for best performing model.
%  Models considered and percent of times chosen as best were following:
%  Adaptive Regularization of Weight Vectors - AROW ($40.4\%$),
%  Logistic Regression ($26.2\%$),
%  Soft Confidence Weighted ($9.5\%$)
%  AdaGradRDA ($8.7\%$),
%  Passive Aggressive ($5.6\%$),
%  Confidence Weighted ($4.9\%$), and
%  Perceptron ($4.7\%$).

% We can notice that standard F measure is only $1\%$  better for
% fully generalized model than for fully specified model, and the difference is even greater ~$7\%$ for population
% weighted F. However the population weighted F measure of $0.6$, and accuracy of %0.67%, indicate moderate performance.
% Trying to apply same level of the domain specificity for all companies does not work as well.
Table \ref{table:strategy_performance} shows the performance of these strategies for the F measure ($F$), accuracy ($A$), and the same metrics weighted by the population of the domain ($F^{W}$ and $A^{W}$).
The full sample population under consideration for each strategy or domain, and the number of samples in the training set, are denoted by $P$ and $T$ respectively.
This same notation is used in Tables \ref{table:strategy_performance} to \ref{table:company_performance}.

We observe from Table \ref{table:strategy_performance} that strategy \textbf{C} leads to a significant $5\%$ increase of F measure over the baselines. 
We observe even higher increases of $15\%$ and $5\%$ for the weighted F
measure and accuracy respectively.
Improving on the best domain selection, by adding the best learning technique in strategy \textbf{D} leads to a further increase of about ~$1\%$ across all metrics.

Figure \ref{fig:prec_rec_2d_distr} visualizes the precision and recall performance for the strategies \textbf{A}, \textbf{C} and \textbf{D} as applied to each domain. We clearly see that the choosing the best domain and best model leads to higher precision and recall values for a wide range of domains.
% Finally we can conclude that baseline strategies using fully specified, or fully generalized model do not perform well
% across wide range of company, language, source variations. This is expected as depending od the company, language, source:
% fully specified domains we may lack training data to build accurate models, and on the other hand for the fully
% generalized models might be over-generalizing as for different domain different keywords and lexicographical
% patterns may indicate actionability. Finally introducing best domain selection, yielded to weighted F measure improvement
% of $15\%$, with additional accuracy uplift of $5\%$.
\begin{table}[ht]
\small
\centering
\setlength\tabcolsep{4pt}
\caption{Strategy Performance Model Breakdown:        % \newline
  \textbf{A} - Logistic model for $\{c,l,s\}$ domain, % \newline % LOGISTIC-FULL\_SPEC,
  \textbf{B} - Logistic model for $\{\}$ domain,      % \newline % LOGISTIC-GENERIC,
  \textbf{C} - Logistic for best domain,              % \newline % LOGISTIC-BEST\_PERMUTED,
  \textbf{D} - Best model-domain                    % \newline % BEST\_MODEL-BEST\_PERMUTED
}
\begin{tabular}{| l || r | r || r | r || r | r | }
  % \hline
  % \rowcolor{Gray} \multicolumn{7}{c}{Multi-column} \\
  \hline
  \rowcolor{Gray} \textbf{S} & \textbf{P} & \textbf{T} & \textbf{$F$} & \textbf{$A$} & \textbf{$F^{W}$} & \textbf{$A^{W}$} \\
  \hline
  \textbf{A}   & 46M	& 1.4M	& 0.701	& 0.701	& 0.558	& 0.687 \\ \hline
  \textbf{B}   & 46M	& 1.4M	& 0.710	& 0.642	& 0.626	& 0.665 \\ \hline
  \textbf{C}   & 46M	& 1.4M	& 0.754	& 0.706	& 0.773	& 0.736 \\ \hline
  \textbf{D}   & 46M	& 1.4M	& \textbf{0.761}	& \textbf{0.719}	& \textbf{0.781}	& \textbf{0.743} \\ \hline
\end{tabular}
\label{table:strategy_performance}
\end{table}

\paragraph{Network Performance}
Next we analyze aggregate performance for Twitter and Facebook where each message is classified using the strategy \textbf{D} described above. 
It is interesting to note that the number of messages created on Twitter is almost 5 times higher than on Facebook, indicating that there is perhaps a greater opportunity for solving the actionability problem on Twitter.
From Table \ref{table:source_performance}, we find that the F measure and accuracy values for Twitter are $0.78$ and $0.75$ respectively, which are $4\%$ and $7\%$ higher than the same metrics for Facebook. Thus we observe that the models perform noticeably better on Twitter.

\begin{table}[ht]
\small
\caption{Network Performance}
\centering
\begin{tabular}{| l || r | r || r | r || r | r | }
  \hline
  \rowcolor{Gray} \textbf{N} & \textbf{P} & \textbf{T} & \textbf{$F$} & \textbf{$A$} & \textbf{$F^{W}$} & \textbf{$A^{W}$} \\
  \hline
  \textbf{tw} & 38M  & 0.8M & 0.78  & 0.75  & 0.79  & 0.75 \\ \hline
  \textbf{fb} &   8M & 0.6M & 0.74  & 0.68  & 0.74  & 0.69 \\ \hline
\end{tabular}
\label{table:source_performance}
\end{table}

\begin{table}[ht]
\small
\setlength\tabcolsep{4pt}
\caption{Language Performance}
\centering
\begin{tabular}{| l || r | r || r | r | r | r | }
  \hline
  \rowcolor{Gray} \textbf{L} & \textbf{P} & \textbf{T} & \textbf{$F$} & \textbf{$A$} & \textbf{$F^{W}$} & \textbf{$A^{W}$} \\
  \hline
  \textbf{en} & 37.6M & 900k & 0.75 & 0.70  & 0.78  & 0.74 \\ \hline
  \textbf{es} & 2.1M  & 90k  & 0.76 & 0.71  & 0.78  & 0.76 \\ \hline
  \textbf{fr} & 1.3M  & 69k  & 0.82 & 0.80  & 0.80  & 0.76 \\ \hline
  \textbf{pt} & 0.8M  & 54k  & 0.72 & 0.68  & 0.74  & 0.71 \\ \hline
  \textbf{tl} & 0.8M  & 47k  & \textbf{0.82} & \textbf{0.80}  & \textbf{0.83} & \textbf{0.79} \\ \hline
  \textbf{id} & 0.4M  & 34k  & 0.74 & 0.68  & 0.78  & 0.74 \\ \hline
  \textbf{af} & 0.3M  & 17k  & 0.78 & 0.76  & 0.79  & 0.75 \\ \hline
  \textbf{it} & 0.4M  & 17k  & 0.78 & 0.76  & 0.78  & 0.72 \\ \hline
  \textbf{nl} & 0.4M  & 34k  & 0.73 & 0.66  & 0.77  & 0.72 \\ \hline
  \textbf{ar} & 0.3M  & 33k  & \textbf{0.82} & \textbf{0.80}  & \textbf{0.82} & \textbf{0.81} \\ \hline
\end{tabular}
\label{table:language_performance}
\end{table}

\begin{table}[htbp]
\small
\setlength\tabcolsep{3pt}
\caption{Company Performance for different Company Profiles}
\centering
\begin{tabular}{| l || r | r || r | r | r | r | }
  \hline
  \rowcolor{Gray} \textbf{Company} & \textbf{P} & \textbf{T} & \textbf{$F$} & \textbf{$A$} & \textbf{$F^{W}$} & \textbf{$A^{W}$} \\
  \hline
  $C_1$ (Media)      & 961k  & 45k   & 0.81  & 0.79  & 0.97  & 0.92 \\ \hline
  $C_2$ (Telco)       & 843k  & 43k   & 0.84  & 0.82  & 0.82  & 0.77 \\ \hline
  $C_3$ (Retailer) & 964k  & 28k   & 0.81  & 0.79  & 0.84  & 0.83 \\ \hline
  $C_4$ (Airline) & 222k  & 16k   & 0.77  & 0.76  & 0.78  & 0.76 \\ \hline
  $C_5$ (Electronics)    & 512k  & 13k   & \textbf{0.90} & \textbf{0.90} & \textbf{0.90} & \textbf{0.90} \\ \hline
  $C_6$ (Beauty Retailer)   & 620k  & 5k    & 0.72  & 0.64  & 0.78  & 0.77 \\ \hline
  $C_7$ (Personal Service) & 158k  & 0.8k  & \textbf{0.93} & \textbf{0.92} & \textbf{0.88} & \textbf{0.87} \\ \hline
  $C_8$ (Manufacturer)    & 19k   & 0.5k  & 0.91  & 0.91  & 0.93  & 0.93 \\ \hline
\end{tabular}
\label{table:company_performance}
\end{table}

\begin{table*}
  \small
  \centering
    \caption{Prediction Examples.}
    \begin{tabular}{ | l | p{15cm} |}
    \hline
    \rowcolor{light-gray} \textbf{Type}  & \textbf{Social Media Message}  \\ \hline \hline
    \multirow{4}{*}{True Positives}          & @YahooCare  Sign in to Yahoo Id gtyij after a long time fails.Sends sec code 2 email id that no longer exists. No option for entering cell no      \\\cline{2-2}
                                  & @Flickr been trying to update payment info \& it will not let me. customer service team is awful. No reply back but have closed account. \\\cline{2-2} \hline \hline
    \multirow{4}{*}{False Positives}          & Blocked 4 Lobster.it accounts on @Flickr. Sick of them trying to 'sell' my work for \$2 whole dollars. \#scam \#photography  \\\cline{2-2}
                                 & @michaelgbaron @crockower if you have comcast or directTV they give you a week free of MLB league pass, just find the channels \\\cline{2-2}
                                 \hline \hline
    \multirow{3}{*}{True Negatives}          & Venezuelan authorities clash with students http://t.co/trvOv21IQr via @Yahoo  \\\cline{2-2}
                                 & @Yahoo: MORE  Two people taken into custody after throwing items over the White House fence: http://t.co/4fPp2bChX6 2d lockdown in a week  \\\cline{2-2} \hline \hline
    \multirow{2}{*}{False Negatives}          & @YahooCare no country listed, url is https://t.co/eKYPvf9msh but based in Ireland.  \\\cline{2-2}
                                 & I was wondering if i can have it reset? I have alot of important emails from school and work, i need help.  \\\cline{2-2}
                                  \hline
    \end{tabular}
  \label{table:examples}
\end{table*}
% , however on the other hand amount of content brands act upon is only 30% greater on Twitter.
% This means both channels are almost equally valuable to the brands in terms of 1:1 engagement,
% however due to public nature of Twitter and amount of the data created Twitter may have higher liability.
% \todo[inline]{Ideally if we have some anecdotal data we can chow here tw and fb have true positive but also false positive.}

\paragraph{Language Performance}
We next compare model performance across a subset of languages.
We see from Table \ref{table:language_performance} that a majority of languages have relatively high F measures, in the range of $0.73$ to $0.82$, with accuracies ranging from $0.66$ to $0.8$. 
The best performing languages are Tagalog and Arabic.
It's interesting to note that though English is the most dominant language in terms of training data, it is not the best-performing one.

We can conclude that the performance of the models is relatively consistent across languages, and is not biased towards the languages with more data. 
This suggests that actionability has a low dependency on the language used, and may instead depend on other factors, such as the consistency of responses from agents. 
Whether features derived from part-of-speech tagging could improve performance, or whether the short-form nature of the messages would limit improvements, remains an open question.
But our analysis here suggests that incorporating other non-linguistic features may provide greater gains.

\paragraph{Company Performance}
Finally, we observe from Table \ref{table:company_performance} that there is a wide variation in the F measures and accuracies for different companies, with the best companies showing more than 90\% F measure and accuracy. 
This variation probably arises from the quality of the ground truth for different companies, which is based upon the human judgement of agents who decide which messages are actionable.
Companies such as those in the telecom sector, with consistently trained staff who respond to customers on defined issues, most likely show better results thanks to better training data.
% \begin{figure*}
% \centering

% \begin{subfigure}[b]{0.48\textwidth}
%   \centering
%   \includegraphics[width=0.95\textwidth]{figure/eps/full_spec_vs_best_eval/network_f1_regular_namespace_vs_best_namespace}
%   \caption{Network original namespace vs best namespace}
%   \label{fig:network_f1_regular_namespace_vs_best_namespace}
% \end{subfigure}
% \begin{subfigure}[b]{0.48\textwidth}
%   \centering
%   \includegraphics[width=0.95\textwidth]{figure/eps/full_spec_vs_best_eval/network_f1_best_namespace_vs_best_model_namespace}
%   \caption{Network best namespace vs model, namespace}
%   \label{fig:network_f1_best_namespace_vs_best_model_namespace}
% \end{subfigure}

% \caption{F1}
% \label{fig:f1_scatter}
% \end{figure*}

% \todo[inline]{Adithya}

%
% Comment out model selection since we can cover it in the Strategy Performance evaluation.
%
% \subsubsection{Compare models}
%
% Comparing approaches from \cite{Crammer09adaptiveregularization} and \cite{Wang12exactsoft}.
%
% \begin{table}[htbp]
% \small
% \setlength\tabcolsep{4pt}
% \caption{Model Selection Comparison}
% \begin{tabular}{|l|p{1cm}|}
%   \hline
%   \rowcolor{Gray} \textbf{Model} & \textbf{Times selected} \\
%   \hline
%   AROW                      & 378 \\ \hline
%   LOGISTIC                  & 245 \\ \hline
%   SOFT CONFIDENCE WEIGHTED  & 89  \\ \hline
%   ADAGRAD RDA               & 82  \\ \hline
%   PASSIVE AGGRESSIVE        & 53  \\ \hline
%   CONFIDENCE WEIGHTED       & 46  \\ \hline
%   PERCEPTRON                & 44  \\ \hline
%   \hline
% \end{tabular}
% \label{table:model_selection}
% \end{table}
%
\paragraph{Examples}
A few examples of actionable and non-actionable messages for different companies are shown in Table \ref{table:examples}.

%% file: texfiles/actionability_conclusion.tex
In this study, we presented a scalable framework for classifying actionable social media messages.
We showed that good results can be achieved even for domains lacking in ground truth data, by leveraging other domains with overlapping attributes.
We built and evaluated models for over 900 domains spanning 75 different companies, 35 languages and 2 social networks. 
Our models were derived using text-based features from lexicons, character markers,
emoticons, readability scores and document length. 
The domain-aware model selection strategies achieved an aggregate population-weighted F measure of 0.78 and accuracy of 0.74, with individual values reaching beyond 0.9 in some cases.

%% file: texfiles/actionability_acknowledgement.tex
We thank our colleagues David Gardner and Kevin Martin for providing insights during this research, and deploying the system to production. 
We also thank Sarah Ellinger for her valuable feedback.

%% file: actionability_ieee.bbl
\begin{thebibliography}{10}
\providecommand{\url}[1]{#1}
\csname url@samestyle\endcsname
\providecommand{\newblock}{\relax}
\providecommand{\bibinfo}[2]{#2}
\providecommand{\BIBentrySTDinterwordspacing}{\spaceskip=0pt\relax}
\providecommand{\BIBentryALTinterwordstretchfactor}{4}
\providecommand{\BIBentryALTinterwordspacing}{\spaceskip=\fontdimen2\font plus
\BIBentryALTinterwordstretchfactor\fontdimen3\font minus
  \fontdimen4\font\relax}
\providecommand{\BIBforeignlanguage}[2]{{%
\expandafter\ifx\csname l@#1\endcsname\relax
\typeout{** WARNING: IEEEtran.bst: No hyphenation pattern has been}%
\typeout{** loaded for the language `#1'. Using the pattern for}%
\typeout{** the default language instead.}%
\else
\language=\csname l@#1\endcsname
\fi
#2}}
\providecommand{\BIBdecl}{\relax}
\BIBdecl

\bibitem{ovum2014white}
A.~Okeleke and S.~Bali, ``{Integrating social media into CRM for next
  generation customer experience},'' Ovum, Tech. Rep., 05 2014.

\bibitem{nemanja-lasta}
N.~Spasojevic, J.~Yan, A.~Rao, and P.~Bhattacharyya, ``Lasta: Large scale topic
  assignment on multiple social networks,'' in \emph{Proc. of ACM Conference on
  Knowledge Discovery and Data Mining (KDD)}, ser. KDD '14, 2014.

\bibitem{Spasojevic:when-to-post}
N.~Spasojevic, Z.~Li, A.~Rao, and P.~Bhattacharyya, ``When-to-post on social
  networks,'' in \emph{Proc. of ACM Conference on Knowledge Discovery and Data
  Mining (KDD)}, ser. KDD '15, 2015.

\bibitem{actionableHaiti}
R.~Munro, ``Subword and spatiotemporal models for identifying actionable
  information in haitian kreyol,'' in \emph{Proceedings of the Fifteenth
  Conference on Computational Natural Language Learning}, 2011, pp. 68--77.

\bibitem{jin2013service}
J.~Jin, X.~Yan, Y.~Yu, and Y.~Li, ``Service failure complaints identification
  in social media: A text classification approach,'' 2013.

\bibitem{agarwal2011sentiment}
A.~Agarwal, B.~Xie, I.~Vovsha, O.~Rambow, and R.~Passonneau, ``Sentiment
  analysis of twitter data,'' in \emph{Proceedings of the Workshop on Languages
  in Social Media}.\hskip 1em plus 0.5em minus 0.4em\relax Association for
  Computational Linguistics, 2011, pp. 30--38.

\bibitem{taboada2011lexicon}
M.~Taboada, J.~Brooke, M.~Tofiloski, K.~Voll, and M.~Stede, ``Lexicon-based
  methods for sentiment analysis,'' \emph{Computational linguistics}, vol.~37,
  no.~2, pp. 267--307, 2011.

\bibitem{glorot2011domain}
X.~Glorot, A.~Bordes, and Y.~Bengio, ``Domain adaptation for large-scale
  sentiment classification: A deep learning approach,'' in \emph{Proceedings of
  the 28th International Conference on Machine Learning (ICML-11)}, 2011, pp.
  513--520.

\bibitem{socher2013recursive}
R.~Socher, A.~Perelygin, J.~Y. Wu, J.~Chuang, C.~D. Manning, A.~Y. Ng, and
  C.~Potts, ``Recursive deep models for semantic compositionality over a
  sentiment treebank,'' in \emph{Proceedings of the conference on empirical
  methods in natural language processing (EMNLP)}, vol. 1631.\hskip 1em plus
  0.5em minus 0.4em\relax Citeseer, 2013, p. 1642.

\bibitem{khan2014tom}
F.~H. Khan, S.~Bashir, and U.~Qamar, ``Tom: Twitter opinion mining framework
  using hybrid classification scheme,'' \emph{Decision Support Systems},
  vol.~57, pp. 245--257, 2014.

\bibitem{Jin:2009:ONM:1557019.1557148}
\BIBentryALTinterwordspacing
W.~Jin, H.~H. Ho, and R.~K. Srihari, ``Opinionminer: A novel machine learning
  system for web opinion mining and extraction,'' in \emph{Proceedings of the
  15th ACM SIGKDD International Conference on Knowledge Discovery and Data
  Mining}, ser. KDD '09.\hskip 1em plus 0.5em minus 0.4em\relax New York, NY,
  USA: ACM, 2009, pp. 1195--1204. [Online]. Available:
  \url{http://doi.acm.org/10.1145/1557019.1557148}
\BIBentrySTDinterwordspacing

\bibitem{nigam2004towards}
K.~Nigam and M.~Hurst, ``Towards a robust metric of opinion,'' in \emph{AAAI
  spring symposium on exploring attitude and affect in text}, 2004, pp.
  598--603.

\bibitem{kanayama2006fully}
H.~Kanayama and T.~Nasukawa, ``Fully automatic lexicon expansion for
  domain-oriented sentiment analysis,'' in \emph{Proceedings of the 2006
  Conference on Empirical Methods in Natural Language Processing}.\hskip 1em
  plus 0.5em minus 0.4em\relax Association for Computational Linguistics, 2006,
  pp. 355--363.

\bibitem{Yui15}
M.~Yui, ``Hivemall,'' 2015, \url{http://mloss.org/software/view/510/}.

\bibitem{bigdata13myui}
M.~Yui and I.~Kojima, ``{A Database-Hadoop Hybrid Approach to Scalable Machine
  Learning},'' in \emph{{Proc. IEEE 2nd International Congress on Big Data}},
  July 2013.

\bibitem{baccianella2010sentiwordnet}
S.~Baccianella, A.~Esuli, and F.~Sebastiani, ``Sentiwordnet 3.0: An enhanced
  lexical resource for sentiment analysis and opinion mining.'' in \emph{LREC},
  vol.~10, 2010, pp. 2200--2204.

\bibitem{dale1948formula}
E.~Dale and J.~S. Chall, ``A formula for predicting readability:
  Instructions,'' \emph{Educational research bulletin}, pp. 37--54, 1948.

\bibitem{kincaid1975derivation}
J.~P. Kincaid, R.~P. Fishburne~Jr, R.~L. Rogers, and B.~S. Chissom,
  ``Derivation of new readability formulas (automated readability index, fog
  count and flesch reading ease formula) for navy enlisted personnel,'' DTIC
  Document, Tech. Rep., 1975.

\bibitem{zhao:2012:mes:2339530.2339772}
\BIBentryALTinterwordspacing
J.~Zhao, L.~Dong, J.~Wu, and K.~Xu, ``Moodlens: An emoticon-based sentiment
  analysis system for chinese tweets,'' in \emph{Proceedings of the 18th ACM
  SIGKDD International Conference on Knowledge Discovery and Data Mining}, ser.
  KDD '12.\hskip 1em plus 0.5em minus 0.4em\relax New York, NY, USA: ACM, 2012,
  pp. 1528--1531. [Online]. Available:
  \url{http://doi.acm.org/10.1145/2339530.2339772}
\BIBentrySTDinterwordspacing

\bibitem{kiritchenko2014sentiment}
S.~Kiritchenko, X.~Zhu, and S.~M. Mohammad, ``Sentiment analysis of short
  informal texts,'' \emph{Journal of Artificial Intelligence Research}, pp.
  723--762, 2014.

\bibitem{Crammer09adaptiveregularization}
K.~Crammer, A.~Kulesza, and M.~Dredze, ``Adaptive regularization of weight
  vectors,'' in \emph{Advances in Neural Information Processing Systems 22},
  2009, pp. 414--422.

\bibitem{Wang12exactsoft}
J.~Wang, P.~Zhao, and S.~C.~H. Hoi, ``Exact soft confidence-weighted
  learning,'' in \emph{In ICML}, 2012.

\end{thebibliography}
